\documentclass[a4paper]{article}
\usepackage[margin=3.4cm]{geometry}
\pdfoutput=1

\usepackage{authblk}

\usepackage{datetime}
\newdateformat{mydateformat}{%
  \THEDAY\ \monthname[\THEMONTH] \THEYEAR}

\usepackage{graphicx}
\graphicspath{{figs/}}

\usepackage{subcaption}
\usepackage{makecell}
\usepackage{threeparttable}

\usepackage{amsmath}
\usepackage{mathtools}
\usepackage{siunitx}  %\num, \qty, \unit

\usepackage{hyperref}

\title{Close contact restriction periods for patients who received iodine-131 therapy for differentiated thyroid cancer}

\author[1,2]{Jake C. Forster}
\author[1]{Daniel Badger}
\author[1,3]{Kevin J. Hickson}
\affil[1]{Medical Physics \& Radiation Safety, South Australia Medical Imaging, Adelaide SA 5000, Australia}
\affil[2]{Department of Physics, School of Physical Sciences, University of Adelaide, Adelaide SA 5005, Australia}
\affil[3]{Allied Health \& Human Performance, University of South Australia, Adelaide SA 5001, Australia}

\date{\mydateformat\today}

\begin{document}

\maketitle

\begin{abstract}

\textit{Objective.} Patients treated with radionuclide therapy may require restrictions on certain activities for a period of time following treatment to optimise protection of the public and ensure the legal dose limit is not exceeded. 
Software may be used to calculate necessary restriction periods for an individual based on longitudinal dose rate measurements from the time of radiopharmaceutical administration. 
A spreadsheet program has been used for this purpose in Australian hospitals for the last two decades. 
However, this spreadsheet has a limitation in that it uses an approximation in the calculation of dose from a contact pattern, which affects the calculated restriction period. 
A computer program called Dorn was developed that provides the same functionality as the spreadsheet but without this approximation. 
Proffered radiation safety advice from Dorn and the spreadsheet were compared. 
\textit{Approach.} Advice from the spreadsheet and Dorn were compared for 55 patients who underwent iodine-131 therapy for differentiated thyroid cancer. 
\textit{Main results.} The restriction periods for caring for infants, close contact with children and sleeping with a partner were typically about 13 hours longer in Dorn than in the spreadsheet, but in some cases were over a week shorter or a month longer.
\textit{Significance.} If the Dorn program is used clinically in place of the spreadsheet, some patients will enjoy shorter restriction periods and the therapy provider can be more confident in their compliance with regulatory requirements and best practice. 
Dorn is freely available from https://doi.org/jg5f.

\end{abstract}

\section{Introduction}

Patients who undergo a diagnostic or therapeutic nuclear medicine (NM) procedure may expose persons who make contact with them to ionising radiation, for example, through work, travel, social, or domestic activities. 
Radiation protection of exposed members of the public should be optimised \cite{rps_c-1}. 
Additionally, in all Australian states and territories, the provider of a NM procedure is legally required to ensure that no member of the public receives an effective dose greater than the prescribed limit as a result of the procedure. 
The effective dose limit for a member of the public is 1~\unit{\milli\sievert} in a year for planned exposure situations, in line with the recommendations of the International Commission on Radiological Protection~\cite{ICRP103}. 
For therapeutic NM procedures in particular, this may necessitate the patient withdraw from certain activities for a period of time following administration of the radiopharmaceutical.

The Australian Radiation Protection and Nuclear Safety Agency has published recommendations for the discharge of NM therapy patients (RPS~4)~\cite{rps4}. 
They emphasise the need for patient-specific radiation safety instructions and expertise from a medical physicist: ``Individualised instructions relevant to the patient's medical and social circumstances should be provided to each patient by the licensed medical specialist responsible for the treatment, in consultation with an experienced medical physicist. The instructions \ldots should be designed to suit the patient's own particular travel and domestic arrangements.''

In situations where patient-specific dose estimates are not available, RPS~4 recommends that, in an effort to meet the 1~\unit{\milli\sievert} public limit, the external ambient dose equivalent rate at 1~\unit{\metre} from the patient should not exceed 25~\unit{\micro\sievert\per\hour} at the time of discharge from hospital. 
RPS~4 also recommends maximum activities of $^{111}$In, $^{131}$I, $^{32}$P, $^{188}$Rh, $^{153}$Sm, $^{89}$Sr, and $^{90}$Y which may be administered in unsealed forms to outpatients, which can be taken as recommended maximum activities retained by the patient at the time of discharge from inpatient therapy.
NM therapy providers may wrongly assume that these discharge criteria ensure compliance with the legal dose limit.
Even if these discharge recommendations are followed, the patient may still require restrictions after being discharged to ensure that no person's dose limit is exceeded. 

J. Cormack and J. Shearer developed software using Microsoft Excel (97-2003) and Visual Basic for Applications called Radionuclide Therapy Close Contact Dose Program (RNTCCDP or ``the spreadsheet'' for short), which can be used to estimate radiation exposures and provide instructions to patients following NM procedures~[personal communication]. 
The spreadsheet has been used clinically in NM departments across Australia for the last two decades.
Briefly, the user enters into the spreadsheet measurements of dose rate at a fixed distance from the patient (either 1, 2, or 3~\unit{\metre}) at multiple time points, starting from the time of administration. 
An exponential or biexponential curve is fitted to the measurements to obtain the whole body clearance function in terms of dose rate.
The instantaneous activity retained by the patient is estimated by multiplying the instantaneous dose rate by the ratio of the administered activity to the initial dose rate (called the baseline method).
Thus, in the spreadsheet, a recommended discharge time is calculated based on when the retained activity reaches the maximum administered activity for outpatient therapy as recommended by RPS~4 (e.g. 600~\unit{\mega\becquerel} for $^{131}$I). If RPS~4 does not contain such a value for the radionuclide being used, the recommended discharge time can be calculated based on when the dose rate at 1~\unit{\metre} reaches 25~\unit{\micro\sievert\per\hour}. 
The clearance function in terms of dose rate can be used to calculate the cumulated dose to a person who shares a particular contact pattern with the patient, resumed after a given period of restriction following administration. The required restriction period is then determined such that the dose to the contacted person is less than the dose constraint. 

RNTCCDP is limited by using an approximation in the calculation of dose from a contact pattern, which affects the calculation of required restriction period. 
A computer program called Dorn (\underline{D}elay \underline{or} \underline{n}ot) was developed that provides the same functionality as the spreadsheet, including a graphical user interface, but implements the full, rigorous calculation of dose from a contact pattern and hence obtains the required restriction periods with greater accuracy than in the spreadsheet. 
To investigate differences in proffered radiation safety advice between the RNTCCDP and Dorn programs, Dorn was applied retrospectively to data recorded during treatments of patients who underwent $^{131}$I therapy for differentiated thyroid cancer~\cite{Silberstein12}, who were managed at the time based on advice from the spreadsheet. 

\section{Methods}

\subsection{Program overview}

\begin{figure}
\begin{center}
\includegraphics[width=8.5cm]{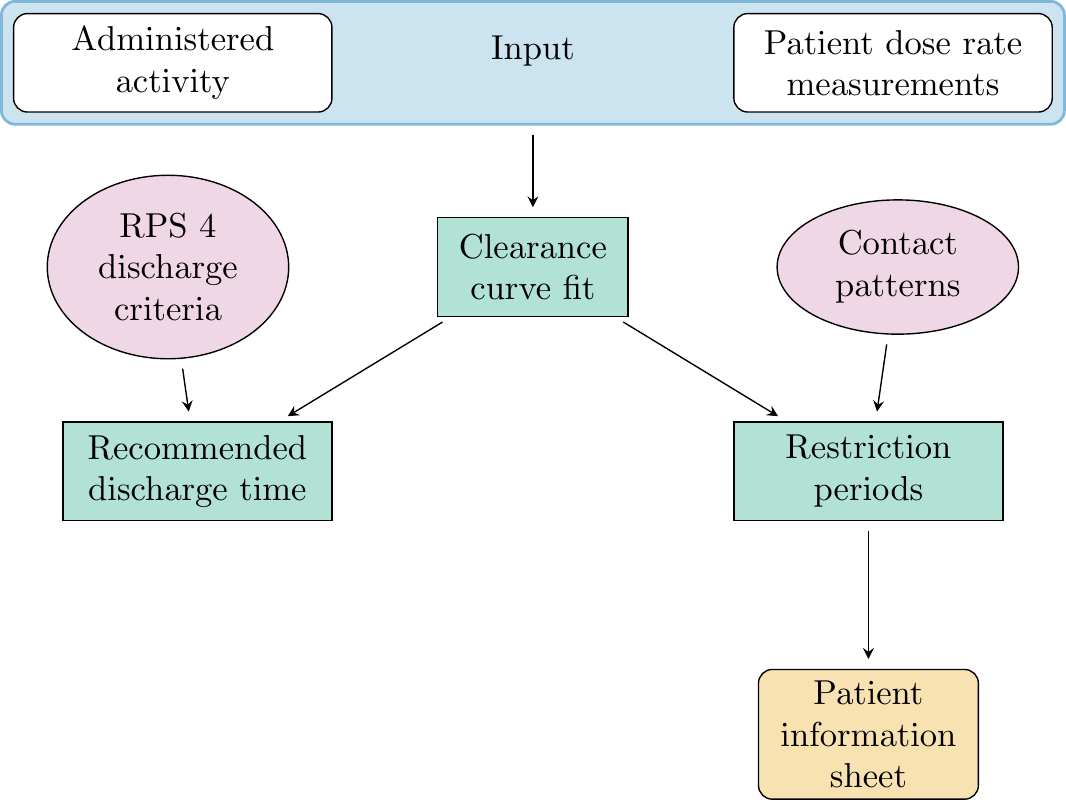}
\caption{\label{fig:flowchart}Overview of the Dorn and RNTCCDP programs, either of which can be used to estimate radiation exposures and provide instructions to patients following NM procedures.}
\end{center}
\end{figure}

The flowchart in Fig.~\ref{fig:flowchart} illustrates the main actions of the Dorn and RNTCCDP programs. 

The RNTCCDP spreadsheet was obtained via File Transfer Protocol. Version 14.0 of the spreadsheet was compared against in this work. 

Dorn was written in Python 3. 
The code was developed on top of well-established Python libraries such as NumPy~\cite{Harris20}, SciPy~\cite{Virtanen20}, pandas~\cite{Mckinney10}, and Matplotlib~\cite{Hunter07}. 
The graphical user interface was created using Python's Tkinter library. 
Data entered by a user is written to and read from XML files. 
A patient information sheet and treatment summary report are generated as Microsoft Word files. 
For the restrictions part of the program, a separate, standalone Python package called glowgreen was written and integrated into Dorn. 
The glowgreen package stores the contact patterns and calculates doses to contacted persons and restriction periods. 

Dorn and glowgreen are open-source software distributed under the MIT License \cite{dorn_v1-9-6, glowgreen_v0-0-4}. 
This work uses version 1.9.6 of Dorn and version 0.0.1 of glowgreen.

\subsection{Dose rate curve fit}
In both the RNTCCDP spreadsheet and Dorn program, an exponential or biexponential curve is fitted to the longitudinal measurements of dose rate at a fixed distance $r$ from the patient, between 1 and 3~\unit{\metre}.
A least-squares fit is performed to an exponential of the form
\begin{equation}
\dot{D}(t) = \dot{D}(0) e^{-t\ln2 / T},  \label{eq:exp}
\end{equation}
which has 2 fit parameters, or to a biexponential
\begin{equation}
\dot{D}(t) = \dot{D}(0) \big[ f_1 e^{-t\ln2 / T_1} + (1-f_1) e^{-t\ln2 / T_2} \big],  \label{eq:biexp}
\end{equation}
with 4 fit parameters.
In the spreadsheet, the Generalised Reduced Gradient nonlinear method was used subject to constraints
\begin{equation}
\dot{D}(0) \geq 0,
\end{equation}
\begin{equation}
0 \leq f_1 \leq 1,
\end{equation}
and
\begin{equation}
\qty{0.36}{\second} \leq T, T_1, T_2 \leq T_p,
\end{equation}
where $T_p$ is the physical half-life of the radionuclide used (8~\unit{\day} for $^{131}$I). 

Dorn uses the curve fit method from the optimize module of the SciPy library~\cite{Virtanen20}, with the Trust Region Reflective algorithm. The initial dose rate is constrained to
\begin{equation}
0 < \dot{D}(0) < A_0 \Gamma \bigg( \frac{\qty{1}{\metre}}{r} \bigg)^{1.5},
\end{equation}
where $A_0$ is the administered activity and $\Gamma$ is the radionuclide's specific gamma or bremsstrahlung dose rate constant at 1~\unit{\metre}. 
The value used for $^{131}$I is \num{7.647e-2}~\unit{\micro\sievert\per\hour\per\mega\becquerel}~\cite{Unger81}. 
The upper bound corresponds to an unattenuated point source, except distance correction from 1~\unit{\metre} to $r$ is performed using an ``inverse 1.5 power'' relationship rather than the inverse square law because it gives a larger upper bound and is more appropriate for an extended patient source~\cite{rps4}. 
The constraint
\begin{equation}
0 < f_1 < 1
\end{equation} 
is applied, and the clearance component half-lives are constrained to 
\begin{equation}
\qty{2.5}{\second} < T < T_p  % rate 1000 /h
\end{equation}
and 
\begin{equation}
\qty{12.5}{\second} < T_1, T_2 < T_p.  % rates 200 /h.
\end{equation}
Values of $T_p$ are taken from the Evaluated Nuclear Structure Data File Database maintained by the National Nuclear Data Center~\cite{NNDC_ENSDF}. For example, $T_p = \num{8.0252}$~\unit{\day} for $^{131}$I. 
The lower bounds assist the algorithm to find optimal parameters. 
Initial guesses of $\dot{D}(0) = (1/2) A_0 \Gamma (\qty{1}{\metre} / r)^{1.5}$, $f_1 = 0.5$ and $T = T_1 = T_2 = T_p / 3$ are provided.

\subsection{Contact patterns}

\begin{figure}
\begin{center}
\includegraphics[width=8.5cm]{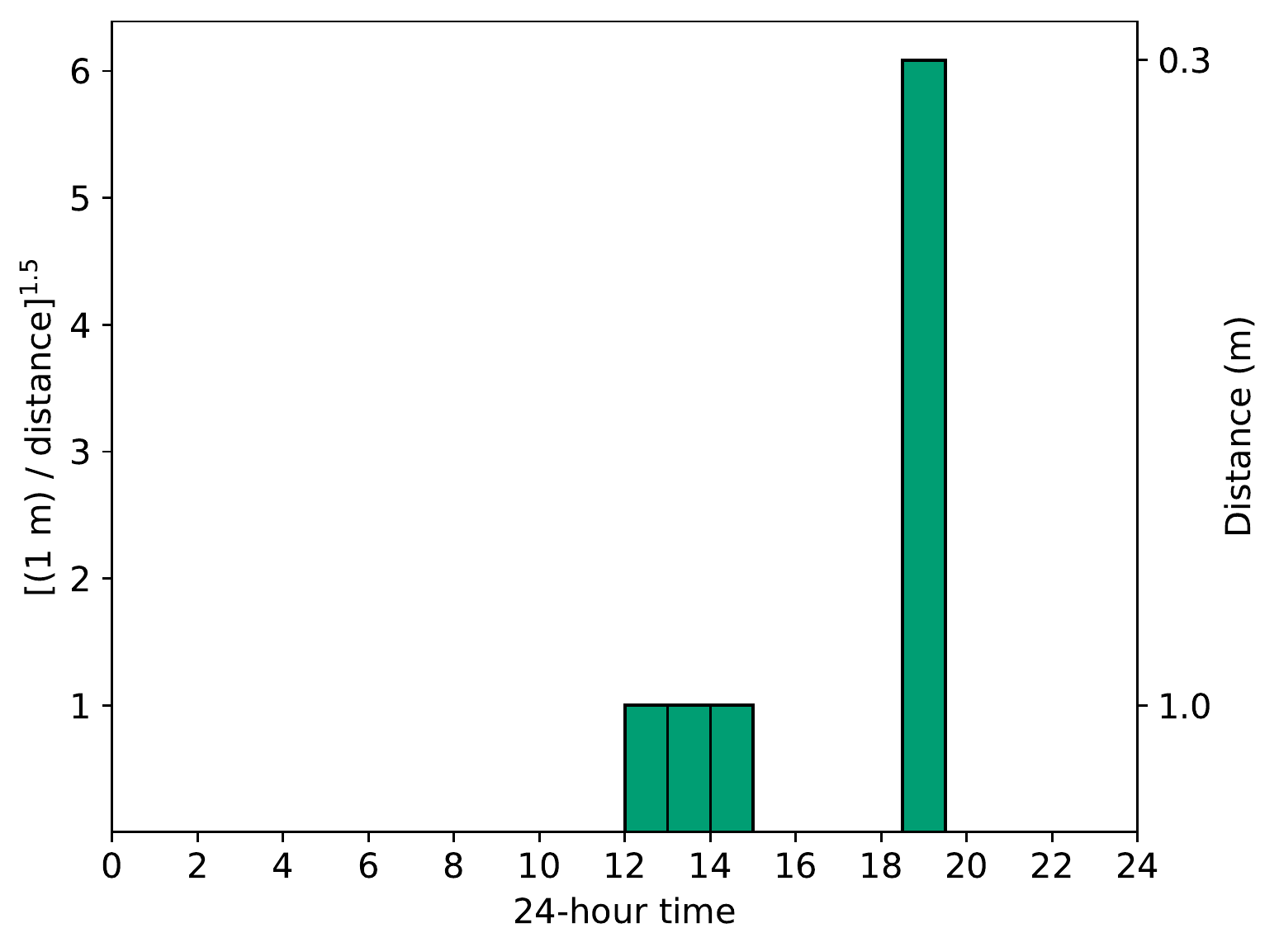}
\caption{\label{fig:cpat}Example of a contact pattern that repeats with a period of 24~\unit{\hour}. 
There is contact every day from noon to 3 PM at a distance of 1~\unit{\metre} and from 6:30 to 7:30 PM at 30~\unit{\centi\metre}. 
The heights of the bars are drawn such that the bar area would be proportional to the dose from the contact if the dose rate was constant.
}
\end{center}
\end{figure}

A contact pattern consisting of $m$ contacts/elements is denoted by $\{\theta_j, c_j, d_j \}$ for $j=1, \ldots, m$, where $\theta_j$ is the time at which the $j$th element starts, $c_j$ is the duration of that element, and $d_j$ is the distance at which it takes place. 
For a contact pattern that repeats with period $p$, $\theta_j$ is defined w.r.t. a reference time, $T_r$. 
For diurnal patterns, midnight is used for the reference time by convention. 
An example of a repeating contact pattern is provided in Fig.~\ref{fig:cpat} for demonstration. 
For a contact pattern that does not repeat, called a once-off contact pattern, $\theta_j$ is instead defined w.r.t. when the contact begins (i.e. $\theta_1 = 0$). 

Cormack and Shearer created 10 unique, repeating, diurnal patterns of close contact for various groups of exposed persons~\cite{Cormack98}, which they implemented in the RNTCCDP spreadsheet. The same contact patterns were implemented in the Dorn program via version 0.0.1 of the glowgreen package. 
They are provided in Supplementary Material for reference.

\subsection{Dose from a contact pattern}

\begin{figure}
\begin{center}
\includegraphics[width=8.5cm]{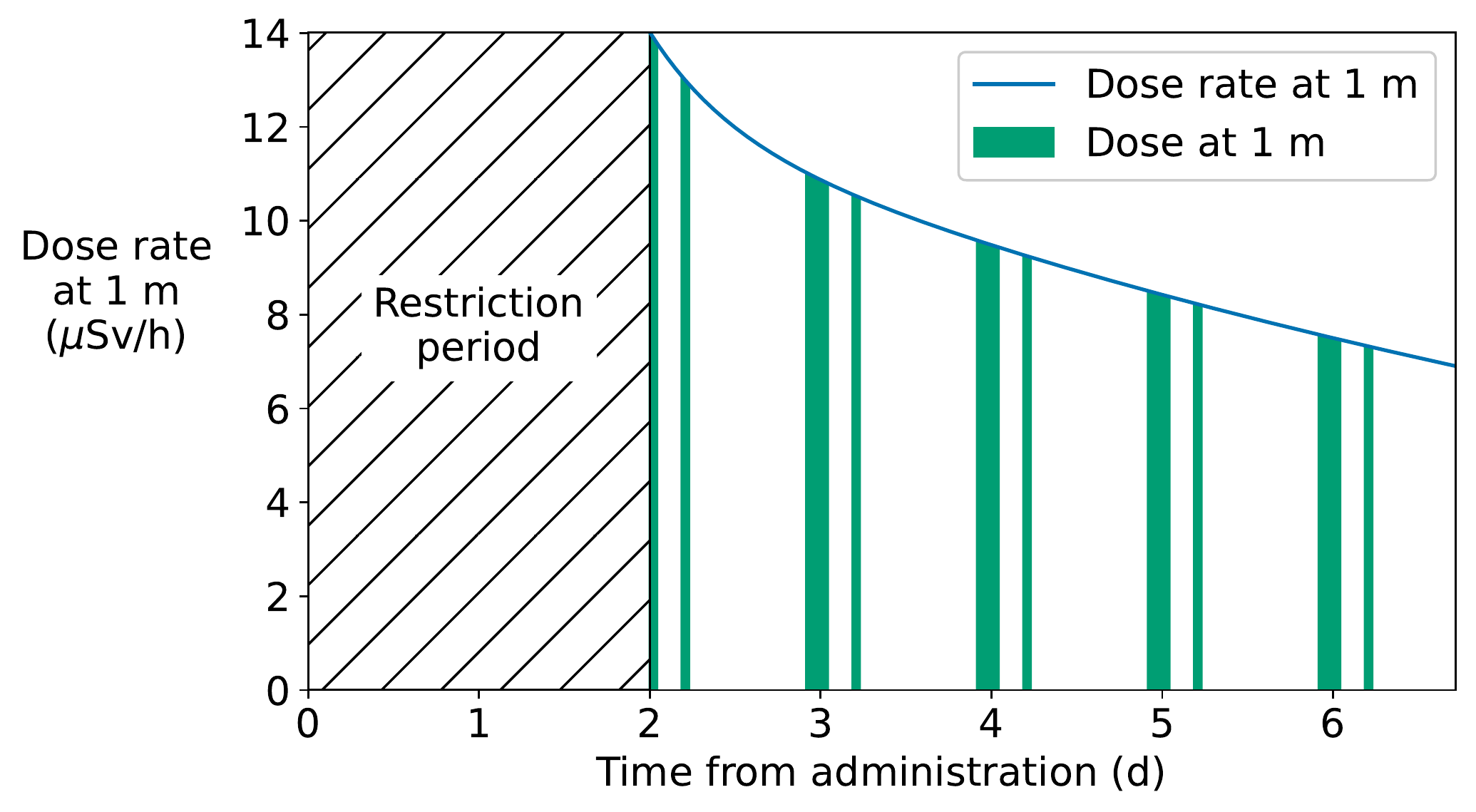}
\caption{\label{fig:dose_example}Example of the dose to a contacted person. Following radiopharmaceutical administration at 2 PM, there is no contact during the restriction period of 48~\unit{\hour} (diagonal hatching), then the repeating contact pattern in Fig.~\ref{fig:cpat} is resumed indefinitely. 
Note the restriction period ends at 2 PM and this pattern includes contact from 2 to 3 PM. 
Each section of shaded area under the curve corresponds to the dose from contact at a distance of 1~\unit{\metre}, and the dose calculation applies a correction factor if the contact occurs at a distance other than 1~\unit{\metre}.
}
\end{center}
\end{figure}

A visual representation of the dose to a contacted person is shown in Fig.~\ref{fig:dose_example}.

The following rigorous expressions for the dose from repeating and once-off contact patterns were implemented in the Dorn program via the glowgreen package. 
Let the dose rate at a distance $r$ from the patient, between 1 and 3~\unit{\metre}, be an n-component exponential:
\begin{equation}
\dot{D}(t) = \dot{D}(0) \sum_{i=1}^n a_i e^{-\lambda_i t}. \label{eq:clearance_function}
\end{equation} 
The curve fit models in Eqs.~\ref{eq:exp} and \ref{eq:biexp} are the $n=1$ and 2 cases, respectively. 
Then the dose for infinite cycles of a repeating contact pattern, commenced after a delay $\tau$ from the administration time $T_a$, where contact may start at any point in the pattern, was shown in Ref.~\cite{Cormack98} to be
\begin{equation}
D(\tau, \infty) = \dot{D}_{1\text{m}} (0) \sum^n_{i=1} \frac{a_i}{\lambda_i} e^{-\lambda_i \tau} \frac{1}{1 - e^{-\lambda_i p}} \sum_{j=1}^m f(d_j) e^{-\lambda_i S_p(\theta_j - \phi)} \big(1-e^{-\lambda_i c_j}\big),
\label{eq:dose_repeating}
\end{equation}
where
\begin{equation}
\dot{D}_{1\text{m}}(0) = \bigg( \frac{r}{\qty{1}{\metre}} \bigg)^{1.5} \dot{D}(0),
\end{equation}
\begin{equation}
f(d_j) = \bigg( \frac{\qty{1}{\metre}}{d_j} \bigg)^{1.5}, \label{eq:distance_factor}
\end{equation}
\begin{equation}
\phi = \big[ \big( T_a - T_r \big) \bmod ( p + \tau ) \big] \bmod p,
\end{equation}
$T_r$ is the reference time following $T_a$; i.e., $T_a - T_r \in [-p, 0[$, and 
\begin{equation}
S_p(x) = 
\begin{cases}
x, & x \geq 0, \\
x + p, & x < 0.
\end{cases}
\end{equation}
Note that if contact starts (i.e. $\tau$ ends) during a pattern element, Eq.~\ref{eq:dose_repeating} does not include the contribution to dose from that pattern element for the first cycle of the pattern. For this reason, in the glowgreen package, elements of repeating patterns with duration greater than 1~\unit{\hour} are broken up into shorter elements, each with duration not exceeding 1~\unit{\hour}. 
Also note that the dose from $t_1$ to $t_2$, each defined from $T_a$, is
\begin{equation}
D(t_1, t_2) = D(t_1, \infty) - D(t_2, \infty). \label{eq:dose_repeating2}
\end{equation}

For a once-off contact pattern, recall $\theta_j$ is defined w.r.t. when $\tau$ ends, and the dose from the pattern is
\begin{equation}
D(\tau) = \dot{D}_{1\text{m}} (0) \sum_{i=1}^n \frac{a_i}{\lambda_i} e^{-\lambda_i \tau} \sum_{j=1}^m f(d_j) e^{-\lambda_i \theta_j} \big( 1 - e^{-\lambda_i c_j} \big).
\label{eq:dose_onceoff}
\end{equation}

\subsubsection{Exposure factor approximation}

In the RNTCCDP spreadsheet, the dose from a repeating contact pattern is approximated as follows. 
First, Cormack and Shearer defined a quantity called the ``exposure factor'' as the ratio of the dose from the contact pattern (Eq.~\ref{eq:dose_repeating},~\ref{eq:dose_repeating2}) to the dose at 1~\unit{\metre} from the patient in the same time interval~[personal communication]. 
They calculated the exposure factor for five different representative clearance functions (thyrotoxic patients, ablation slow clearance, ablation fast clearance, euthyroid patients and ``normal'' clearance rate) using administration at 10~AM and contact starting at 24~\unit{\hour} and continuing forever.
The average value of the five exposure factors was taken as a generic exposure factor, $k$. 
The generic exposure factors used in the spreadsheet are listed in Table~\ref{tab:exposure_factors}.

Then, in the spreadsheet, the dose from a repeating contact pattern from $t_1$ to $t_2$, each defined from the time of administration, was approximated by
\begin{equation}
D^*(t_1, t_2) = k D_{1\text{m}}(t_1, t_2), \label{eq:dose_approx}
\end{equation}
where $D_{1\text{m}}(t_1, t_2)$ is the dose at 1~\unit{\metre} from the patient from $t_1$ to $t_2$, using the patient's clearance function.
To emphasise, Eq.~\ref{eq:dose_approx} is only an approximation of the dose from a contact pattern (Eqs.~\ref{eq:dose_repeating},~\ref{eq:dose_repeating2}); the true exposure factor for a patient depends on the time of administration, $t_1$, $t_2$, and the patient's clearance function.

\begin{table}
\begin{center}
\caption{\label{tab:exposure_factors}Generic exposure factor values used in the RNTCCDP spreadsheet.}
\begin{tabular}{p{0.6\textwidth} c}
\hline
Contact pattern & Exposure factor \\
\hline
Caring for infants (normal) & 1.75 \\
Caring for infants (demanding or sick) & 5.37 \\
Close contact with 2-5 year old children & 2.89 \\
Close contact with 5-15 year old children & 1.29 \\
Sleeping with spouse or partner & 4.67 \\
Close contact with adult friends and family & 1.01 \\
Close contact with informed persons caring for patient & 1.28 \\
Daily public transport to and from work & 0.66 \\
Return to work involving prolonged close contact with others & 0.68 \\
Return to work not involving prolonged close contact with others & 0.37 \\
\hline
\end{tabular}
\end{center}
\end{table}

\subsection{Restriction periods}

The restriction period is the time (or least time, up to the temporal resolution of the pattern elements) from administration to when a contact pattern begins or is resumed, such that a person who shares that contact pattern with the patient receives a lifetime dose equal to (or less than) a given dose constraint. 
From Eq.~\ref{eq:dose_approx}, the dose from an infinitely repeating contact pattern, resumed after a delay $\tau$ from administration, is approximated in the RNTCCDP spreadsheet by
\begin{equation}
D^{*}(\tau, \infty) = k D_{1\text{m}}(\tau, \infty). \label{eq:dose_approx_inf}
\end{equation}
The restriction period, $\tau_r$, required to achieve a dose constraint, $D_c$, is then found by solving
\begin{equation}
k D_{1\text{m}}(\tau_r, \infty) = D_c. \label{eq:required_delay_approx}
\end{equation}
Eq.~\ref{eq:required_delay_approx} is solved in the RNTCCDP spreadsheet using Newton's method.
Newton's method for finding a solution of $f(x) = 0$ with initial guess $x = x_1$ is
\begin{equation}
x_{n+1} = x_n - \frac{f(x_n)}{f^{\prime}(x_n)}.
\end{equation}
In this case, a zero is sought for 
\begin{align}
f(\tau) &= k D_{1\text{m}}(\tau, \infty) - D_c \\
& = k \int_{\tau}^{\infty} \mathrm{d}t \dot{D}_{1\text{m}}(t) - D_c.
\end{align}
For a clearance function that is n-component exponential 
\begin{equation}
\dot{D}_{1\text{m}}(t) = \dot{D}_{1\text{m}}(0) \sum_{i=1}^n a_i e^{-\lambda_i t},
\end{equation}
we have 
\begin{equation}
f(\tau) = k \dot{D}_{1\text{m}}(0) \sum_{i=1}^n \frac{a_i}{\lambda_i} e^{-\lambda_i \tau} - D_c 
\end{equation}
and
\begin{equation}
f^{\prime}(\tau) = - k \dot{D}_{1\text{m}}(0) \sum_{i=1}^n a_i e^{-\lambda_i \tau}.
\end{equation}
Therefore Newton's method is
\begin{align}
\tau_{n+1} &= \tau_n - \frac{f(\tau_n)}{f^{\prime}(\tau_n)} \\
&= \tau_n + \frac{\sum_{i=1}^n \frac{a_i}{\lambda_i} e^{-\lambda_i \tau_n} - \frac{D_c}{k \dot{D}_{1\text{m}}(0)}}{\sum_{i=1}^n a_i e^{-\lambda_i \tau_n}}.
\end{align}
The restriction period is calculated in the spreadsheet using iteration (circular reference), stopping after 100~iterations or earlier if the change is less than 0.001~\unit{\hour}, with initial guess $\tau_1 = 0$. 
Calculated restriction periods less than 0 are set to 0.

In contrast, in the Dorn program, the dose from infinite cycles of a repeating contact pattern, resumed after a delay $\tau$ from administration, is calculated rigorously using Eq.~\ref{eq:dose_repeating}. 
To determine the required restriction period, the dose is calculated for contact resuming on the hour following administration, then on subsequent hours until the dose is less than the dose constraint. 
If the end of the restriction period does not coincide with the start of a pattern element, the restriction period is extended to the start of the next pattern element. That way, it is clear that contact can be resumed at the end of the restriction period. 

Once-off contact patterns can also be used in the Dorn program. 
The dose from a once-off contact pattern is calculated using Eq.~\ref{eq:dose_onceoff}. 
If the clearance function is exponential, the restriction period is solved exactly by
\begin{equation}
\tau_r = \frac{1}{\lambda} \ln \bigg( \frac{\dot{D}_{1\text{m}}(0) \rho}{D_c \lambda} \bigg), \label{eq:restriction_period_onceoff_exp}
\end{equation}
where
\begin{equation}
\rho = \sum_{j=1}^m f(d_j) e^{-\lambda \theta_j} \bigg( 1 - e^{-\lambda c_j} \bigg).
\end{equation}
For biexponential clearance, the restriction period is found by numerically solving Eq.~\ref{eq:dose_onceoff} for the dose constraint using the fsolve function in the optimize module of the SciPy library~\cite{Virtanen20}, with a starting estimate of zero. 

Table~\ref{tab:restrictions} lists the contact pattern and dose constraint pairings for which restriction periods were calculated in both the spreadsheet and Dorn. 
The dose constraint was set to 1~\unit{\milli\sievert} except for exposure to radiosensitive material (0.1~\unit{\milli\sievert}) and appropriately informed carers who knowingly and willingly provide comfort and support to the patient (5~\unit{\milli\sievert}). The latter constraint is in line with RPS~4, which recommends a dose constraint of 5~\unit{\milli\sievert} per treatment episode for this exposure group.

\begin{table}
\begin{center}
\begin{threeparttable}
\caption{\label{tab:restrictions}Contacts for which restriction periods were calculated.}
\begin{tabular}{l p{0.6\textwidth} c}
\hline
Label & Contact pattern & \makecell{Dose constraint\\(\unit{\milli\sievert})} \\
\hline
A & Caring for infants (normal) & 1 \\
B & Caring for infants (demanding or sick) & 1 \\
C & Close contact with 2-5 year old children & 1 \\
D & Close contact with 5-15 year old children & 1 \\
E & Sleeping with spouse or partner & 1 \\
F & Sleeping with informed person supporting patient\tnote{a} & 5 \\
G & Close contact with adult friends and family & 1\\
H & Close contact with informed persons caring for patient & 5 \\
I & Daily public transport to and from work & 1\\
J & Return to work involving prolonged close contact with others & 1\\
K & Return to work not involving prolonged close contact with others & 1\\
L & Work with radiosensitive materials\tnote{b} & 0.1 \\
\hline
\end{tabular}
\begin{tablenotes}
\item [a] Same contact pattern as E.
\item [b] Same contact pattern as K.
\end{tablenotes}
\end{threeparttable}
\end{center}
\end{table}

In both programs, if a patient is expected to receive more than one cycle of a radionuclide therapy in a single year, 
the dose constraints for contacts with members of the public and radiosensitive materials are reduced by a factor of the number of cycles in a year. 

\subsection{Software comparison study}

To compare results between the RNTCCDP spreadsheet and the Dorn program, Dorn was applied retrospectively to data from a cohort of patients who received $^{131}$I sodium iodide for treatment of differentiated thyroid cancer and whose subsequent radiation safety advice was based on information generated by the spreadsheet. 
The study included patients who were treated at The Queen Elizabeth Hospital in Woodville South, South Australia or at Lyell McEwin Hospital, Elizabeth Vale, South Australia, between January 2019 and April 2021, whose $^{131}$I was administered orally with 1 or 2 capsules, and who received recombinant human thyroid stimulating hormone (rhTSH; Thyrogen) in preparation.
At least one of the authors was involved in the primary care of each patient.

Patient dose rates were measured at 2~\unit{\metre} using the same calibrated detector. 
Patients were included whose dose rates were measured with either a Ludlum Model 2241 digital scaler-ratemeter (Ludlum Measurements, Inc) with Ludlum Model 44-9 Geiger-M{\"{u}}ller (GM) pancake-type detector and ambient dose equivalent filter for energy compensation; or a Radiation Alert Ranger (S.E. International, Inc), which is an uncompensated GM tube. 
For each patient's clearance function, the same curve fit model (either exponential or biexponential) was used in Dorn as was originally used in the spreadsheet.

Not all types of restrictions apply to a single patient, and patients were only presented with restrictions relevant to their lives. 
But for the purpose of the comparison study, restriction periods were calculated for all contacts in Table~\ref{tab:restrictions} for all patients.

\section{Results}

A total of 69 thyroid cancer patients treated with $^{131}$I met the inclusion criteria for the software comparison study. 
However, 3 of these patients were excluded due to an issue with the spreadsheet that prevented viewing of entered data. 
A further 11 patients were excluded due to suspected errors in the data, such as the dose rate measurement distance was entered as 1~\unit{\metre} or the detector calibration factor was not applied to the dose rate measurements. 
This left 55 patients in the study. 
The median time from administration to discharge from hospital was 24~\unit{\hour}, and the patient's dose rate was measured a median of 6 times before discharge (Table~\ref{tab:demographic}).

\begin{table}
\begin{center}
\caption{\label{tab:demographic}Demographic and management of radioiodine patients in the software comparison study.}
\begin{tabular}{p{0.4\textwidth} c c}
\hline
Characteristic & No. (\%) & Median (min--max) \\
\hline
Sex & & \\
\quad Male & 10 (18) & \\
\quad Female & 45 (82) & \\
Age - yr & & 48 (19--84) \\
$^{131}$I administered activity -~\unit{\giga\becquerel} & & 4.1 (2.0--5.9) \\
Time from administration to discharge -~\unit{\hour} & & 24 (21--45) \\
Number of dose rate measurements acquired & & 6 (4--8) \\
Number of dose rate measurements used in curve fit & & 6 (3--8) \\
Dose rate curve fit model & & \\
\quad Exponential & 3 (5) & \\
\quad Biexponential & 52 (95) & \\
\hline
\end{tabular}
\end{center}
\end{table}

The curve fitted in the Dorn program to the longitudinal patient dose rate measurements had a smaller root-mean-square deviation than the curve fitted in the RNTCCDP spreadsheet for all but one patient, and the difference ranged from \num{-2.6} to 0.07~\unit{\micro\sievert\per\hour}. 
There was one patient whose biexponential curve fit in both the spreadsheet and Dorn included a component with a half-life close to the upper bound. 
Another three patients had such a component in the curve fit from Dorn which was not present in the curve fit from the spreadsheet.

A recommended discharge time was calculated based on a dose rate of 25~\unit{\micro\sievert\per\hour} at 1~\unit{\metre}, or a retained activity of 600~\unit{\mega\becquerel}; whichever was later. 
Then the recommended discharge times from Dorn and the spreadsheet differed by at most 2.8~\unit{\hour}, with a median difference of 10~\unit{\minute}.

\begin{figure}
\begin{center}
\includegraphics[width=8.5cm]{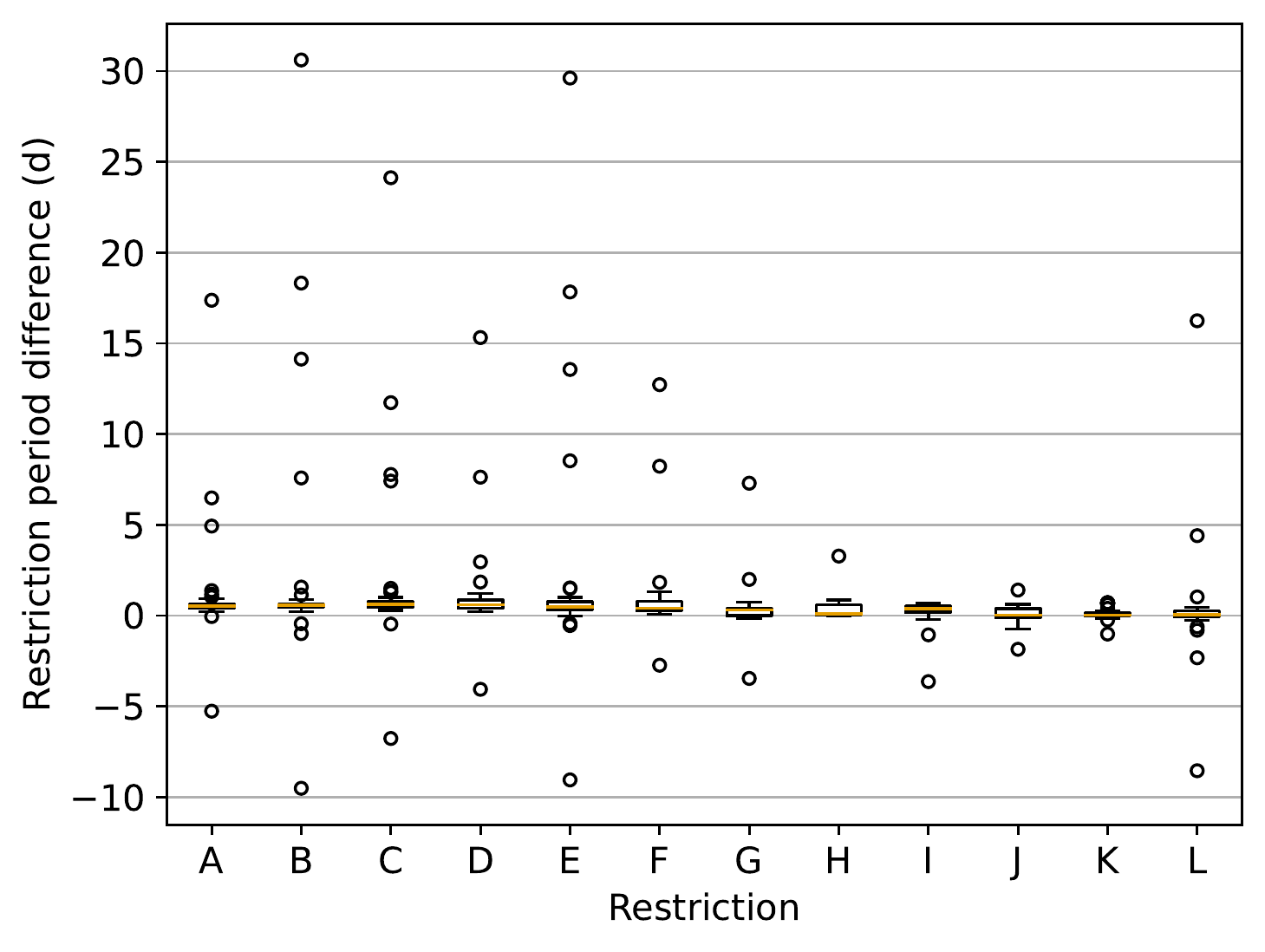}
\caption{\label{fig:restriction_period_difference}Box plots of the difference between the restriction periods in Dorn and RNTCCDP for the same patient. Restrictions are labelled A--L as in Table~\ref{tab:restrictions}. 
The median differences were: A, 12.8~\unit{\hour}; B, 13.5~\unit{\hour}; C, 14.8~\unit{\hour}; D, 14.4~\unit{\hour}; E, 11.6~\unit{\hour}; F, 10.0~\unit{\hour}; G, 7.3~\unit{\hour}; H, 2.1~\unit{\hour}; I, 9.4~\unit{\hour}; J, 0.7~\unit{\hour}; K, 0.4~\unit{\hour}; L, 1.3~\unit{\hour}. 
}
\end{center}
\end{figure}

\begin{figure}
\begin{center}
\includegraphics[width=8.5cm]{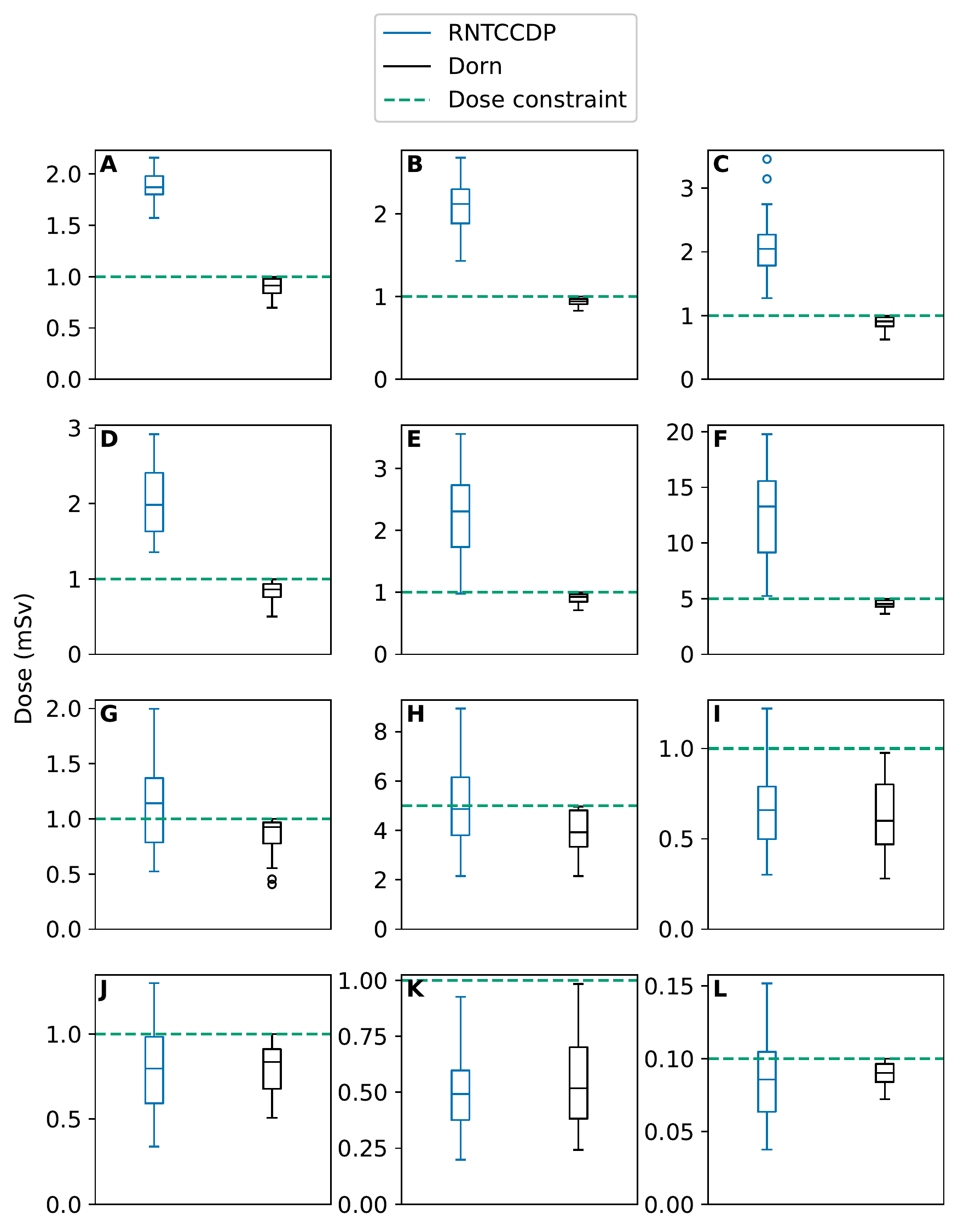}
\caption{\label{fig:dose}Doses to contacted persons with the restriction periods in RNTCCDP (left) and Dorn (right), as calculated by the rigorous method (Eq.~\ref{eq:dose_repeating}). The dose from the RNTCCDP spreadsheet is the dose from infinite cycles of the repeating contact pattern, resumed after the restriction period in the spreadsheet, using the curve fit in the spreadsheet. Similarly for the dose from Dorn. Restrictions are labelled A--L as in Table~\ref{tab:restrictions}. 
}
\end{center}
\end{figure}

The restriction periods in both the spreadsheet and Dorn were often extended when the patient had a biexponential clearance function that contained a component with the physical half-life. 
Restriction periods differed between the spreadsheet and Dorn due to their different curve fits and calculation methods. 
For contacts relating to caring for infants and children and sleeping with another person, the restriction period was typically about 13~\unit{\hour} longer in Dorn than in the spreadsheet (Fig.~\ref{fig:restriction_period_difference}). 
This suggests that the approximation used in the spreadsheet for the dose from an infinitely repeating contact pattern (Eq.~\ref{eq:dose_approx_inf}), using the exposure factor values provided, may have underestimated the dose to the contacted person in some cases. 
This was investigated by performing the rigorous calculation of dose (Eq.~\ref{eq:dose_repeating}) for the restriction periods advised in the spreadsheet and using the curve fits obtained in the spreadsheet. 
Indeed, the restriction periods calculated in the spreadsheet frequently resulted in doses above the constraints for several of the theoretical contact patterns considered (Fig.~\ref{fig:dose}).

\begin{figure}
\begin{center}
\includegraphics[width=8.5cm]{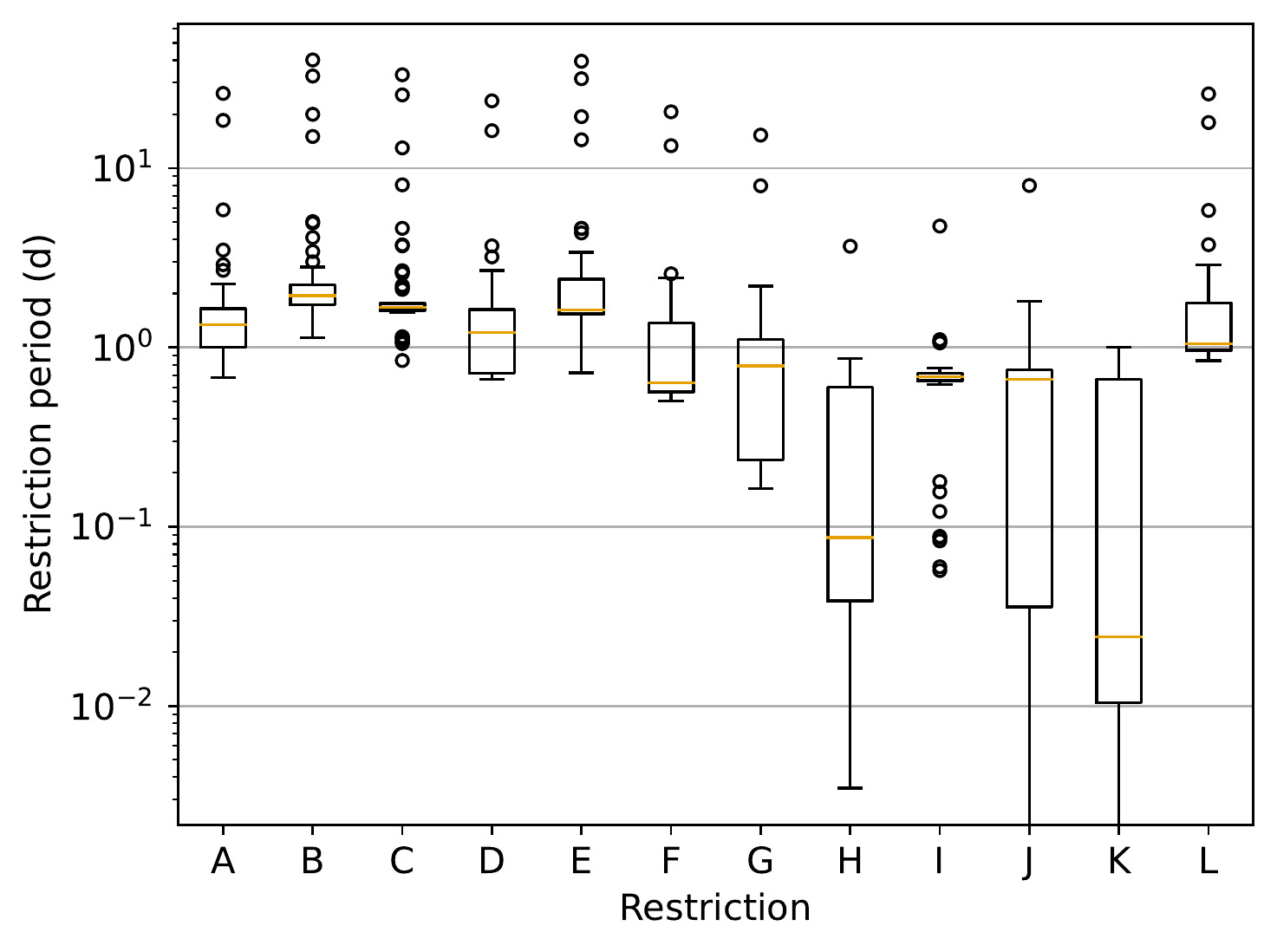}
\caption{\label{fig:restriction_period}Restriction periods from the Dorn program for the cohort of thyroid cancer patients treated with $^{131}$I. Restrictions are labelled A--L as in Table~\ref{tab:restrictions}. 
The median restriction periods were: A, 32.1~\unit{\hour}; B, 46.8~\unit{\hour}; C, 40.1~\unit{\hour}; D, 29.1~\unit{\hour}; E, 38.9~\unit{\hour}; F, 15.2~\unit{\hour}; G, 18.9~\unit{\hour}; H, 2.1~\unit{\hour}; I, 16.5~\unit{\hour}; J, 15.9~\unit{\hour}; K, 0.6~\unit{\hour}; L, 25.1~\unit{\hour}. 
}
\end{center}
\end{figure}

The restriction periods generated by Dorn for this patient cohort are shown in Fig.~\ref{fig:restriction_period}.

\section{Discussion}

While the provision of patient-specific restrictions following radionuclide therapy should be considered best practice, there are several potential pitfalls of which to be aware.
First, the fitted clearance function may be inaccurate when projected outside the period of actual observations, and inaccuracy in the clearance function may impact the calculated restriction periods.
For example, a biexponential fit that contains a component with a half-life close to the upper bound is likely to overestimate the dose rate (and retained activity) at later times, which may extend the calculated restriction period. 

The curve fit will more accurately represent the underlying clearance function if the patient empties their bladder before each dose rate measurement, thereby reducing sawtooth effects. 
It should also be noted that the first dose rate measurement in the morning may be elevated due to a reduction in the patient's metabolic rate during sleep~\cite{Sharma10}, which could cause the clearance rate to be underestimated. 
But the most obvious improvement to the curve fit can be made by providing additional dose rate measurements extending into the second day after administration and beyond. 

This work applied the published contact patterns from Ref.~\cite{Cormack98}, but they may not be representative of the actual contacts had by a patient, which are expected to vary considerably between patients. 
Since the 1~\unit{\milli\sievert} public limit is a legal requirement, unless there is specific information about the contacts a patient will have, the ``worst-case'' scenario must be applied even though it is not appropriate for most patients. 

There are problems with advising restriction periods that are longer than necessary; for example, due to using a contact pattern that represents the worst-case scenario: The patient is inconvenienced, or they ignore the restrictions, or medical practitioners think the restrictions are too long and modify them or advise patients to ignore them. 
Therefore, an important area for development is to improve the validity of the contact patterns and provide alternatives to the worst-case scenario. 

The Dorn and spreadsheet programs can be used to estimate the radiation exposure from close contact with the patient, but neither takes into account radiation exposure from contamination. 
$^{131}$I therapy for thyroid cancer is delivered in the inpatient setting in part due to the risk of contamination, particularly during the first 24~\unit{\hour} following administration. 
But after about 24~\unit{\hour}, which is when the patients in this study were usually discharged, the activity available to be released in contamination events is greatly reduced.

\subsection{Software comparison}

The software comparison study found the restriction periods were considerably different when the dose was calculated rigorously instead of using the exposure factor approximation.
This approximation, which allows the use of Newton's method to rapidly calculate the required restriction period, was appealing in the context of a spreadsheet program with the computational resources available twenty years ago. 
With modern computing power, the full dose calculation can be performed for more accurate restriction periods. 

There was a suspected error in the data entered into the spreadsheet program for several patients reviewed, which suggests the spreadsheet may be difficult to use correctly. 
User error is expected to be less common in Dorn because it provides a simple, linear workflow and additional quality control checks on the data entered. 

As with the spreadsheet, the Dorn program is not limited to $^{131}$I therapy for thyroid cancer; it has the potential to be used for any radionuclide therapy, provided longitudinal patient dose rate measurements are acquired.

\section{Conclusion}

Software such as the RNTCCDP spreadsheet or the Dorn program can be used to calculate required restriction periods based on measurements of the patient's dose rate at multiple time points. 
Dorn calculates the required restriction periods in a more accurate way than the spreadsheet. 
Using dose rate data from patients who received $^{131}$I for remnant thyroid cancer ablation, the restriction periods for caring for infants, close contact with children, and sleeping with a partner were typically 13~\unit{\hour} longer in Dorn than in the spreadsheet, but in some cases were over a week shorter or a month longer. 
If Dorn is used clinically in place of the spreadsheet, some patients will enjoy shorter restriction periods and the therapy provider can be more confident in their compliance with regulatory requirements and best practice.

\end{document}

\maketitle

\section*{Contact patterns}
The theoretical contact patterns used in this work from Ref.~\cite{Cormack98} are shown in Figs.~S1--S10.

\makeatletter
\@for\cpat:={cpat_Caring_for_infants_normal,cpat_Caring_for_infants_demanding_or_sick,cpat_Close_contact_with_2-5_year_old_children,cpat_Close_contact_with_5-15_year_old_children,cpat_Sleeping_with_spouse_or_partner,cpat_Close_contact_with_adult_friends_and_family,cpat_Close_contact_with_informed_persons_caring_for_patient,cpat_Daily_public_transport_to_and_from_work,cpat_Return_to_work_involving_prolonged_close_contact_with_others,cpat_Return_to_work_not_involving_prolonged_close_contact_with_others}\do{

\begin{figure}
\begin{center}
\includegraphics[width=8.5cm]{\cpat}
\caption{}
\end{center}
\end{figure}

}